\documentclass[preprint]{elsarticle}

\biboptions{sort&compress}
\usepackage{epsfig}
\usepackage{longtable}
\usepackage{amsmath}
\usepackage{enumitem}

\journal{Physics Letters B}

\begin{document}

\begin{frontmatter}
\title{Darkening the Little Higgs}
\author[loc:triumf]{Travis A. W. Martin\fnref{em:tam}}
\author[loc:triumf]{Alejandro de la Puente\fnref{em:adlp}}

\fntext[em:tam]{Email: tmartin@triumf.ca}
\fntext[em:adlp]{Email: adelapue@triumf.ca}

\address[loc:triumf]{TRIUMF, 4004 Wesbrook Mall, Vancouver, BC V6T 2A3, Canada}

\begin{abstract}

We present a novel new method for incorporating dark matter into little Higgs models in a way that can be applied to many existing models without introducing $T$-parity, while simultaneously alleviating precision constraints arising from heavy gauge bosons. The low energy scalar potential of these dark little Higgs models is similar to, and can draw upon existing phenomenological studies of, inert doublet models. Furthermore, we apply this method to modify the littlest Higgs model to create the next to littlest Higgs model, and describe details of the dark matter candidate and its contribution to the relic density.

\end{abstract}
\end{frontmatter}

\textbf{Introduction}. -- Little Higgs (LH) models~\cite{ArkaniHamed:2001nc,ArkaniHamed:2002pa,ArkaniHamed:2002qx} are extensions of the Standard Model (SM) that stabilize the electroweak scale with a light Higgs boson and weakly coupled new physics. These models resolve the fine-tuning problem within the SM by embedding the Higgs boson within a non-linear sigma field, and by introducing new gauge and fermion states that result in a collective breaking of the scalar Higgs potential. This collective symmetry breaking ensures cancellation of the quadratic divergences that result from radiative corrections from gauge boson and top quark loops that plague the SM Higgs boson.

The challenges in constructing a modern little Higgs model include: generating a natural mass hierarchy between the heavy top partner(s) and heavy gauge bosons that fits within precision electroweak constraints; avoiding the generation of a dangerous singlet in the scalar potential~\cite{Schmaltz:2008vd}; and, in light of the mounting evidence for dark matter, the inclusion of a dark matter candidate. For example, the littlest Higgs model~\cite{ArkaniHamed:2002qy,Han:2003wu} and simplest little Higgs model~\cite{Schmaltz:2004de} do not include a dark matter candidate, and are largely constrained by precision measurements~\cite{Hewett:2002px,Csaki:2003si,Reuter:2012sd}. While the bestest little Higgs (BLH) model~\cite{Schmaltz:2010ac} resolves these precision constraint issues by including a custodial $SU(2)$ symmetry and introducing a second non-linear sigma field that couples only to the gauge bosons, it does not include a dark matter candidate. 

It has been noted that certain classes of little Higgs models may contain discrete symmetries that can be used to introduce a viable dark matter candidate. In particular, three such classes of models have been studied: theory space models~\cite{BirkedalHansen:2003mpa}, $T$-parity models~\cite{Cheng:2004yc} and skyrmion models~\cite{Murayama:2009nj,Gillioz:2010mr}. In the latter, $T$-parity~\cite{Cheng:2004yc} requires new fermions, forces the gauge couplings to be equal, $g_{1}^{(\prime)}=g_{2}^{(\prime)}$, forces conservation of a $T$-charge for all interactions (therefore, the lightest $T$-odd state is stable), and results in an elimination of the triplet vacuum expectation value (vev). Theory space models~\cite{BirkedalHansen:2003mpa} contain a $Z_{4}$ symmetry that can be used to interchange the non-linear sigma model fields amongst themselves. Within this class of models, the scalar identified with the SM-like Higgs boson breaks the $Z_{4}$ symmetry down to a $Z_{2}$ symmetry after electroweak symmetry breaking (EWSB), and the lightest particle charged under the $Z_{2}$ may become a viable dark matter candidate. Additionally, dark matter can arise in some little Higgs models from topological considerations~\cite{Murayama:2009nj,Gillioz:2010mr}. In these models, skyrmions take the form of topological solitons.

In this letter, we explore an alternative method of introducing dark matter to little Higgs models by incorporating a second non-linear sigma field, $\Delta$. This expands upon the concept introduced in the bestest little Higgs model~\cite{Schmaltz:2010ac} and in another T-parity model~\cite{Pappadopulo:2010jx}, and provides a relatively simple means of implementing an inert doublet potential~\cite{Deshpande:1977rw,LopezHonorez:2006gr,Dolle:2009fn} - in effect, we prescribe a means of little Higgs-ing the inert doublet models. It should be noted that this is not the only implementation of an inert doublet potential in little Higgs models (see \cite{Brown:2010ke}). This presents a new class of little Higgs models, dark little Higgs (DLH) models, which follow the general structure: 
\begin{itemize}[itemsep=0mm]
\item{duplicate global symmetry ($G_\Delta/H_\Delta$ duplicates group structure of $G_\Sigma/H_\Sigma$) that breaks at scale $F > f$;}
\item{$G_\Delta$ gauged in the same way as $G_\Sigma$;}
\item{and, fermions transform only under $G_\Sigma$.}
\end{itemize}
Since fermions do not transform under the second global symmetry, the complex doublet embedded in $\Delta$ does not develop a non-zero vev, and thus remains as a possible dark matter candidate. Additionally, by following this prescription, the heavy top partner masses are disconnected from the mass of the heavy gauge bosons, which relaxes electroweak precision constraints on the models without reintroducing fine-tuning constraints.

In this letter, we describe the details of a simplistic version of this by modifying the littlest Higgs model into the next to littlest Higgs model, a DLH class model, and explore the relic abundance generated by the inert doublet.\\

\textbf{The Model}. -- The littlest Higgs is based on a non-linear sigma field ($\Sigma$) that parametrizes an $SU(5)_{\Sigma}/SO(5)_{\Sigma}$ coset space. We introduce a second non-linear sigma field, $\Delta$, parametrizing a separate coset space, $SU(5)_{\Delta}/SO(5)_{\Delta}$, but require that both the $SU(5)_{\Sigma}$ and $SU(5)_{\Delta}$ global symmetries contain the same gauged $[SU(2) \times U(1)]^2$ subgroup. Fermions transform only under the $SO(5)_\Sigma$ symmetry, and so the scalar doublet embedded in $\Delta$ does not acquire a radiatively generated negative mass squared. As with other little Higgs models, this description does not explain the physics origin of the non-linear sigma model, which is relevant only at or above the ``compositeness" scale $\lambda \sim 4 \pi f$.

The $SU(5)_\Sigma$ symmetry is broken to $SO(5)_\Sigma$ at a scale $f$, as in the littlest Higgs, while $SU(5)_\Delta$ is broken to $SO(5)_\Delta$ at a scale $F$ ($> f$).  The vacuum expectation values that generate this breaking are the same as in the littlest Higgs model, given by:
\begin{eqnarray}
\Sigma_0 = \left(\begin{array}{ccc}0 & 0 & 1\!\!1_{2\times2} \\0 & 1 & 0 \\1\!\! 1_{2\times2} & 0 & 0\end{array}\right), & \Delta_0 = \left(\begin{array}{ccc}0 & 0 & 1\!\!1_{2\times2} \\0 & 1 & 0 \\1\!\! 1_{2\times2} & 0 & 0\end{array}\right).
\end{eqnarray}
The non-linear sigma fields are then parameterized as:
\begin{eqnarray}
\Sigma(x) = e^{2i\Pi_\Sigma/f} \Sigma_0, \;\; &\;\; \Delta(x) = e^{2i\Pi_\Delta/F} \Delta_0
\end{eqnarray}
where $\Pi_\Sigma = \sum_a \pi_\Sigma^a X^a $ and $\Pi_\Delta =  \sum_a \pi_\Delta^a X^a$, summing over the 14 Goldstone bosons ($\pi^a_{\Sigma,\Delta}$) corresponding to the 14 generators ($X^a$) in each sector. In the littlest Higgs model, four fields corresponding to four of the broken generators are eaten to give mass to the heavy gauge bosons, and three are eaten to give mass to the SM gauge bosons, leaving seven observable scalar states. In our model, there are 14 broken generators for each of the $\Sigma$ and $\Delta$ sectors (total of 28), and a total of seven are eaten to give mass to the gauge bosons, leaving 21 observable scalars.

Both $SU(5)$ symmetries are gauged by the same $[SU(2) \times U(1)]^2$ subgroups, with generators $Y_1=\mathrm{diag}(-3,-3,2,2,2)/10$ and $Y_2 = \mathrm{diag}(-2,-2,-2,3,3)/10$ for the two $U(1)$ groups, and 
\begin{eqnarray}
Q_1^a = \left(\begin{array}{ccc}\sigma^a/2 & 0 & 0 \\0 & 0 & 0 \\0 & 0 & 0\end{array}\right) & Q_2^a = \left(\begin{array}{ccc} 0 & 0 & 0 \\0 & 0 & 0 \\0 & 0 & -\sigma^{a*}/2\end{array}\right)
\end{eqnarray}
for the two $SU(2)$ groups. In this notation, $\sigma^a$ are the Pauli matrices.

The new fields of the $\Delta$ non-linear sigma field are embedded in the Pion matrix as:
\begin{eqnarray}
\Pi_\Sigma = \left(\begin{array}{ccc}0 & h^\dagger/\sqrt{2} & \phi^\dagger \\ \xi/\sqrt{2} & 0 & h^*/\sqrt{2} \\\phi & h^T/\sqrt{2} & 0\end{array}\right)  \!\!\!\!& \!\!\!\!&+ (Q_1^a - Q_2^a)\eta^a  \cr
&& + \sqrt{5}(Y_1-Y_2)\sigma \cr\cr
\Pi_\Delta = \left(\begin{array}{ccc}0 & \xi^\dagger/\sqrt{2} & \chi^\dagger \\ \xi/\sqrt{2} & 0 & \xi^*/\sqrt{2} \\\chi & \xi^T/\sqrt{2} & 0\end{array}\right)  \!\!\!\!& \!\!\!\!&+ (Q_1^a - Q_2^a)\alpha^a  \cr
&& + \sqrt{5}(Y_1-Y_2)\beta
\end{eqnarray}
where $\xi$ and $\chi$ are the analogous fields to the $h$ and $\phi$ from the $\Sigma$ sector, and the real triplet ($\eta^a$, $\alpha^a$) and singlet ($\sigma$, $\beta$) representations of the two non-linear sigma fields mix to form a combination that becomes the longitudinal components of the heavy gauge bosons ($\alpha_{e}^a = (f \eta^a + F\alpha^a)/\sqrt{f^2+F^2}$ and $\beta_e = (f \sigma + F\beta)/\sqrt{f^2+F^2}$), and an orthogonal combination that is physical. 

These new fields couple to the gauge bosons in the normal way, via the kinetic term of the Lagrangian, such that,
\begin{equation}
\mathcal{L}_K = \frac{f^2}{8}Tr[(D_\mu\Sigma)(D^\mu\Sigma)^\dagger]+ \frac{F^2}{8}Tr[(D_\mu\Delta)(D^\mu\Delta)^\dagger].
\end{equation}
The covariant derivative is given as
\begin{eqnarray}
D_\mu \Sigma(\Delta) &=&\partial_\mu \Sigma(\Delta) - i \sum_j g_j W_j^a (Q_j^a \Sigma(\Delta) + \Sigma(\Delta) Q_j^{aT})\cr
&& - i \sum_j g_j^\prime B_j(Y_j \Sigma(\Delta) + \Sigma(\Delta)Y_i),
\end{eqnarray}
where the sum is over $j=1,2$ for each of the two $SU(2)\times U(1)$. The heavy gauge boson masses pick up an extra contribution proportional to $F^2$, such that $M_{W_H}^2 = \frac{1}{4}(g_1^2+g_2^2)(f^2+F^2)$ and $M_{B_H}^2 = \frac{1}{20}(g_1^{\prime 2} + g_2^{\prime 2})(f^2+F^2)$.

The Coleman-Weinberg (CW) derived couplings ($\lambda$'s) for the $h$ and $\phi$ in the scalar potential remain predominantly unchanged at leading order, as factors of $F$ cancel out, leaving a dependence only on the scale $\Lambda$. Factors of $F$ still contribute in the $\mu^2$ term, which contains logarithmic divergences, through the masses of the heavy gauge bosons. The negative contribution from the heavy quark sector is still dominant in the $\mu^2$ term in the potential, and induces spontaneous symmetry breaking. 

We can examine the degree of fine tuning in the model as in ~\cite{Casas:2005ev} by examining the logarithmically divergent contributions to the $\mu^2$ term in the scalar potential. Examining $\delta_T \mu^2$, $\delta_W \mu^2$, $\delta_B \mu^2$ and $\delta_\phi \mu^2$, we similarly find that $\delta_T \mu^2$ is responsible for the largest degree of fine tuning of the $\mu$ parameter. For a Higgs boson mass of 125 GeV, and scale parameters $f=1$~TeV and $F=5$~TeV, we find $\delta_W \mu^2 / m_h^2 < 11$, as compared with $\delta_T \mu^2 / m_h^2 < 180$. Thus it is clear that the degree of fine tuning in the model is controlled by the heavy quark sector, and larger values of $M_{W^\prime}$ that result in a relaxation of electroweak (EW) precision constraints are viable without significantly increasing the degree of fine tuning.

Other EW precision constraints arise in the model as a result of the triplet vev, $v^\prime$. The scalar potential for $\phi$ is unchanged from the littlest Higgs model, which provides the relation $v^\prime < (v/4f) v$~\cite{Csaki:2002qg}. Since the $v^\prime$ contributions to the EW precision observables are subdominant over those proportional to $v^2/f^2$ (or $M_W^2/M_{W^\prime}^2$)~\cite{Csaki:2002qg} for most of the parameter space, the overall constraints on the scales $f$ and $F$ arising from EW precision observables will be improved over the original littlest Higgs model. In~\cite{Han:2003wu}, it was argued that $v^\prime$ passes the constraints on $\Delta g_1^Z$ for values of $v^\prime < 10\% v$, which is easily satisfied within the NLH model.

The masses of the $\phi$ and $h$ fields in the $\Sigma$ sector are similar to those found in the littlest Higgs. The $\chi$ triplet obtains a quadratically divergent mass from the one loop CW potential, while the $\xi$ doublet only obtains a logarithmically divergent mass. The dominant terms in the masses of these states are given by:
\begin{eqnarray}
M_{\chi}^2 &=& \frac{3}{16\pi^2} \frac{\Lambda^2}{f^2+F^2}(M_{W_H}^2+M_{B_H}^2)\\
M_{\xi}^2 &=& \frac{3}{128\pi^2} (f^2+F^2)g_1^2g_2^2 \log\left(\frac{\Lambda^2}{M_{W_H}^2}\right)\cr
&&+\frac{3}{1280\pi^2} (f^2+F^2)g_1^{\prime 2}g_2^{\prime 2} \log\left(\frac{\Lambda^2}{M_{B_H}^2}\right)
\end{eqnarray}
After electroweak symmetry breaking, contributions to the mass of the $\xi$ doublet proportional to the square of the vev, $v^2$, arising from the one loop logarithmic terms create a small mass splitting between the neutral and charged states, and the neutral component becomes the lightest state.

The CW potential also generates a small negative mass for the $\sigma$ field, which would necessarily induce spontaneous symmetry breaking. To avoid this, we introduce a small, positive, explicit mass term for the $\sigma$ field of the form:
\begin{eqnarray}
V_{\Delta} = \lambda_{\Delta}F^4 Tr[T_\Delta (\Delta-\Delta_0) T_\Delta (\Delta-\Delta_0)^\dagger]
\end{eqnarray}
The operator $T_\Delta$ has a certain amount of flexibility, so long as it does not violate gauge invariance. We take $T_{\Delta}=\mathrm{Diag}[0,0,1,0,0]$, which is a minimal solution that avoids contributions to the masses of other fields. The value of $\lambda_{\Delta}$ is restricted by perturbativity constraints only, but is taken to be $O(10^{-1})$. The masses of the real singlet and triplet are then given by:
\begin{eqnarray}
M_{\sigma}^2 &\approx& \frac{16 f^2 F^2}{5(f^2 + F^2)} \lambda_\Delta - \frac{f^2 F^2}{40 \pi^2 (f^2 + F^2)} \frac{\lambda_t^4}{s_t^2 c_t^2} \log\left( \frac{\Lambda^2}{M_T^2}\right)\cr
M_{\eta_0}^2 &\approx& \frac{3 (f^2+F^2) g_1^2 g_2^2}{128 \pi^2} \log\left(\frac{\Lambda^2}{M_{W_H}^2}\right)\cr
M_{\eta^\pm}^2 &\approx& \frac{9 (f^2+F^2) g_1^2 g_2^2}{128 \pi^2} \log\left(\frac{\Lambda^2}{M_{W_H}^2}\right)
\end{eqnarray}

The parameters in the model are thus limited to the two symmetry breaking scales, $f$ and $F$; three mixing angles ($s=\sin\theta_g=g/g_1$, $s^\prime=\sin\theta_g^\prime=g^\prime/g_1^\prime$, $s_t=\sin\theta_t=\lambda_t/\lambda_1$) identical to those defined in the littlest Higgs model; the explicit scalar coupling $\lambda_\Delta$; and two parameters which characterize the higher scale physics that is responsible for cancelling the divergences in the the Coleman-Weinberg potential, $a$ and $a^\prime$. The parameters $a$ and $a^\prime$ are the same as those defined in the littlest Higgs model; for a more detailed discussion, see~\cite{Han:2003wu,Casas:2005ev}.

The heavy gauge bosons ($W_H$, $Z_H$, and $A_H$) and complex scalar triplet states ($\phi$ and $\chi$) are typically quite heavy in the NLH model, at least in the several TeV range, while the heavy top partner can easily be lighter than a TeV. The real scalar triplet ($\eta$) and singlet states ($\sigma$) masses typically vary between a few hundred GeV and the low TeV range. The mass of the complex doublet, $\xi$, typically takes values below 1 TeV, which will be discussed in the following section.\\

\textbf{Dark Matter}. -- As defined, the $\xi$ is a degenerate, two component dark matter candidate (scalar and pseudoscalar), with a mass in the $O(100~\mathrm{GeV})$ range. Such degenerate, complex DM candidates necessarily generate a large direct detection signal through a vector coupling to nuclei~\cite{TuckerSmith:2001hy}. A recent study of this phenomenon is present in \cite{Earl:2013jsa}. This can be resolved by introducing an explicit symmetry breaking term into the Lagrangian that breaks the accidental symmetry that maintains the mass degeneracy between the $Re[\xi]$ and $Im[\xi]$ fields, such as: 
\begin{eqnarray}
V_{\Sigma\Delta} = -\lambda_{\Sigma\Delta} f^2 F^2 Tr[T_{\Sigma\Delta} (\Sigma-\Sigma_0) T_{\Sigma\Delta} (\Delta-\Delta_0)^\dagger]\cr+h.c.
\end{eqnarray}
The operator $T_{\Sigma\Delta}$ has a certain amount of flexibility, but must be chosen to prevent mixing between the $\xi$ and $h$ fields. An operator constructed from a linear combination of the generators of the $[SU(2)\times U(1)]^2$ symmetry is well motivated for this, since these operators already preserve gauge invariance \footnote{A similar argument motivates the operator $T_{\Delta}$ from the previous section.}. In particular, $T_{\Sigma\Delta} = n_1 \mathrm{Diag}[1,1,0,0,0] + n_2 \mathrm{Diag}[0,0,0,1,1]$ will resolve the mass splitting, and respects a $Z_2$ symmetry that protects the $\xi$ field from decaying. The following contribution to the masses of the neutral $\xi$ states are generated with such an operator:
\begin{eqnarray}
\delta m_{\xi_0\equiv Re[\xi]}&=&-(n_1+n_2)^2 \lambda_{\Sigma\Delta} v^2 - 4 (n_1^2 - n_2^2) f v^\prime\cr\cr
\delta m_{\Xi_0\equiv Im[\xi]}&=&(n_1-n_2)^2 \lambda_{\Sigma\Delta} v^2 + 4 (n_1^2 - n_2^2) f v^\prime
\end{eqnarray}
As described in \cite{TuckerSmith:2001hy}, the mass splitting needs only to be of $O(100~\text{keV})$, indicating that small values of $\lambda_{\Sigma\Delta}$ are acceptable. Of note, a value of $\lambda_{\Sigma\Delta} \sim 0.01$ will produce a mass splitting of $O(1~\mathrm{GeV})$. Additional contributions to the masses of the $\chi_0$, $\eta_0$, $\sigma_0$, $\chi^\pm$ and $\eta^\pm$ will also arise, but can be eliminated by setting either $n_1=0$ or $n_2=0$. However, we use $n_1=n_2=1$, as a simple implementation, since the contribution to the masses of the other states will be of a similarly small nature, and thus unimportant to the overall phenomenology of the model.

One interesting aspect of the NLH model is that, after decoupling the electroweak scalar triplets, the heavy top and the heavy electroweak gauge bosons, the scalar potential reduces to the inert doublet potential:
\begin{eqnarray}
V&=&\mu^{2}_{1}|H_{1}|^{2}+\mu^{2}_{2}|H_{2}|^{2}+\lambda_{1}|H_{1}|^{4}+\lambda_{2}|H_{2}|^{4}  \\
&+&\lambda_{3}|H_{1}|^{2}|H_{2}|^{2}+\lambda_{4}|H^{\dagger}_{1}H_{2}|^{2}+\lambda_{5}Re[(H^{\dagger}_{1}H^{2})^{2}]\nonumber.
\end{eqnarray}
Within this framework, $H_1\equiv h$ is the SM Higgs doublet that is spontaneously broken; $H_2\equiv \xi$ is the doublet from the $\Delta$ sector; and the $\lambda_{\Sigma\Delta}$ terms contribute to $\lambda_4$ and $\lambda_5$, where $\lambda_4 v^2$ contributions to the neutral component of $H_2$ preserve a mass degeneracy and $\lambda_5 v^2$ contributions generate a mass splitting between the scalar and pseudoscalar states.

Many of the studies of inert doublet dark matter can thus be applied to the NLH model. For example, by comparing supersymmetry search results from LEPI and LEPII to the inert doublet model, the authors in~\cite{Lundstrom:2008ai} found mass constraints of:
\begin{eqnarray}
m_{\xi_0}&\ge& 80~\text{GeV} \nonumber \\
m_{\Xi_0}&\ge& 100~\text{GeV} \nonumber \\
m_{\Xi_0}-m_{\xi_0}&\le& 8~\text{GeV}.\label{eq:LEPbounds}
\end{eqnarray}
These constraints are easily satisfied within the NLH model.

\begin{figure}[htb]
   \begin{center}
	\includegraphics[width=0.8\textwidth]{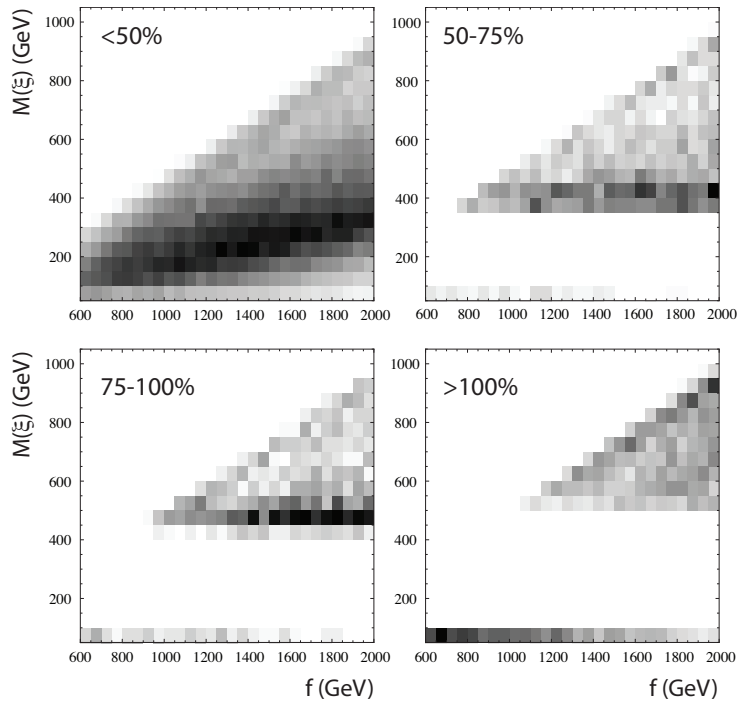}
   \end{center}
   \vspace{-4mm}
   \caption{Density plot of the $\sim130k$ parameter points examined, where darker shaded regions correspond to a greater number of parameter points resulting in the value of $m_\xi$ for a given value of $f$ that account for the given relic abundance relative to the measured value. Histogram density scales are unique to each subplot.}
\label{fig:MC}
\end{figure}

The Planck collaboration~\cite{Ade:2013lta} has recently published updated results on the relative relic abundance of dark matter, giving a best fit value of $\Omega h^2 = 0.11889$. Using FeynRules~\cite{Degrande:2011ua}, we have implemented the model in the software package MicrOMEGAs~\cite{Belanger:2010gh} and calculated the relic density arising from the lightest stable state in our model, assuming a Higgs boson mass of 125~GeV. Using a Monte Carlo method to select parameter values ($0.05 < s < 0.95$, $0.05 < s^\prime < 0.95$, $600 < f < 2000$, $0.2 < f/F < 0.8$, $0.05 < s_t < 0.95$, $0 < \lambda_{\Sigma\Delta} < 0.5$, $0 < \lambda_\Delta < 0.5$, $a=1$, $a^\prime=1$), $\sim130k$ models were generated and the relic density calculated. Of the models that produced viable masses for the DM candidate, $65.4\%$ of the parameter sets could only account for less than half of the relic density. Only $1.2\%$ of the parameter space explored could account for $75-100\%$ of the relic density, while $2.6\%$ of the parameter space resulted in a relic density larger than the Planck result. The distribution of these results can be seen in Fig. \ref{fig:MC}, where the medium (50-100 GeV) and heavy ($>500$ GeV) dark matter regions for inert doublet models, as described in~\cite{Gustafsson:2010zz}, are clear in the $75-100\%$ plot.

\begin{figure}[htb]
   \begin{center}
	\includegraphics[width=0.8\textwidth]{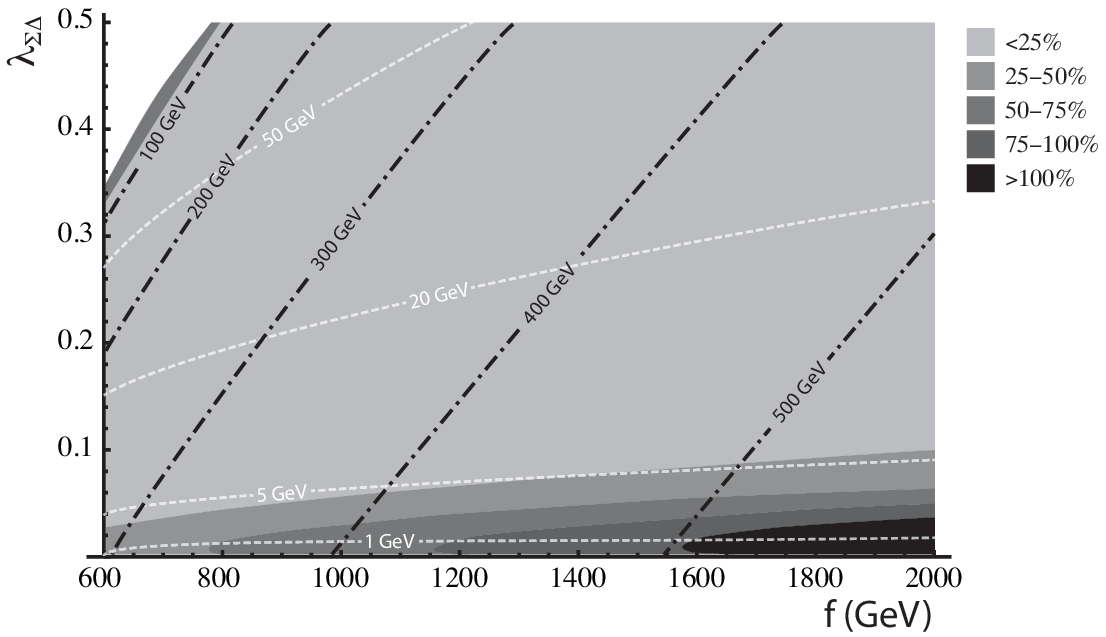}
   \end{center}
   \vspace{-4mm}
   \caption{Contour showing the relative relic abundance in the next to littlest Higgs model, using fixed other parameters while varying $f$ and $\lambda_{\Sigma\Delta}$. Percentage values give the ratio of $\Omega h^2_{\mathrm{NLH}}/\Omega h^2_{\mathrm{Planck}}$. The dark region in the upper left corner indicates a region where the mass of the $\xi_0 \rightarrow 0$ rapidly, and a finer scanning is needed to fully understand the relic abundance. The white region in the upper left corner indicates the region where $m_{\xi_0}^2 < 0$, and thus not viable. Black dot-dashed lines show the mass of the dark matter candidate, while white dashed lines show the mass separation between the $\xi$ and $\Xi$ particles.}
\label{fig:contour}
\end{figure}

To directly examine the effect of the $\lambda_{\Sigma\Delta}$ and $f$ dependence of the results, the other parameters were fixed ($s/c = 0.25$, $s^\prime/c^\prime = 0.25$, $F=3000$, $\lambda_\Delta = 0.2$, $a=1$, $a^\prime=1$) while $f$ and $\lambda_{\Sigma\Delta}$ were varied over the range $600 < f < 2000$, $0 <\lambda_{\Sigma\Delta} < 0.5$. The results from this scan are given in Fig. \ref{fig:contour}. While a large region of the parameter space results in a DM candidate that can only account for $< 25\%$ of the relic abundance, there still exists substantial parameter space where the $\xi_0$ can account for the full relic abundance.

The parameter $\lambda_{\Sigma\Delta}$ plays two roles in this, which accounts for the contours observed in Fig.~\ref{fig:contour}. The first is that $\lambda_{\Sigma\Delta}$ directly controls the coupling between $\xi$ and $h$: increasing the value of $\lambda_{\Sigma\Delta}$ results in a larger $\xi\xi \rightarrow hh$ annihilation rate. This is the dominant annihilation mode in the region where the relic abundance predicted in the model is in the $<25\%$ range. The second role is that $\lambda_{\Sigma\Delta}$ controls the mass separation between $\xi$ and $\Xi$ ($\Delta M = \sqrt{M_\xi^2 + \lambda_{\Sigma\Delta} v^2} - M_\xi$), which affects the co-annihilation rates~\cite{Krawczyk:2013wya}. The relic abundance is otherwise understood as a manifestation of the inert doublet models ~\cite{LopezHonorez:2006gr,Dolle:2009fn,Lundstrom:2008ai,Gustafsson:2010zz}.

\textbf{Summary}. -- We have presented a new class of little Higgs models, called dark little Higgs models, that employ the little Higgs method of resolving the large quadratically divergent Higgs boson mass present in the standard model, and generate an inert doublet model that can simultaneously account for dark matter. In addition, we have presented a simple implementation of this class of models in the form of the next to littlest Higgs model - a modification of the littlest Higgs model -  and explored the relic abundance predicted in the model. We found that a heavy dark matter candidate with a mass on the order of 500~GeV can be generated for regions of the parameter space that can account for the observed relic abundance, in agreement with existing studies of inert doublet models.

\textbf{Acknowledgements}. -- The authors would like to thank Heather Logan, Thomas Gr\'egoire, and David Morrissey for guidance and assistance. This work was supported by the National Science and Engineering Research Council of Canada (NSERC).

\bibliographystyle{model1-num-names}

\end{document}